# Asymmetric imaging: Reciprocity bounds and manifestations

Romil Audhkhasi[1], Anna Wirth-Singh[1], Rose Johnson[1], Maksym Zhelyeznyakov[1], Vladimir Yarmolik[2], Rafael Piestun[3,4], Andrea Alù[5,6], Arka Majumdar[1,2,*] and Owen Miller[7,8,*]
[1]Department of Electrical and Computer Engineering, University of Washington, Seattle, Washington 98195, United States
[2]Department of Physics, University of Washington, Seattle, Washington 98195, United States
[3]Department of Electrical, Computer and Energy Engineering, University of Colorado, Boulder, Colorado 80309, United States
[4]Department of Physics, University of Colorado, Boulder, Colorado 80309, United States
[5]Photonics Initiative, Advanced Science Research Center, City University of New York, New York, NY 10031, United States
[6]Physics Program, Graduate Center, City University of New York, New York, NY 10016, United States
[7]Department of Applied Physics, Yale University, New Haven, Connecticut 06520, United States
[8]Department of Electrical and Computer Engineering, Yale University, New Haven, Connecticut 06520, United States
[*]Corresponding authors: arka@uw.edu, owen.miller@yale.edu

**Abstract**
Recent demonstrations of unidirectional imaging using diffractive multilayers have brought a concept akin to optical isolation to imaging, yet the fundamental limits of such asymmetry have remained unknown. Here we develop a general theory of asymmetric imaging in linear, passive, reciprocal optical systems and derive a fundamental bound on the achievable unidirectionality. Reciprocity requires the forward and reverse transmission operators to have identical singular values, a wave-optical analogue of étendue equality, with a simple consequence: perfect asymmetry is attainable only for sets of $M \leq N/2$ orthogonal object patterns, where $N$ is the number of available object/image modes; for $M > N/2$, the average power asymmetry is bounded by $(N - M)/M$. Guided by this result, we present two complementary realizations: refractive imaging systems formed by bulk lenses and stops that attain the half-occupation limit, and a 1 cm-diameter all-silicon metasurface doublet that exhibits asymmetric focusing at 4 µm while fully complying with Lorentz reciprocity. We fabricate the doublet by direct laser writing and experimentally demonstrate its focusing asymmetry. By establishing mechanism-independent limits, our results provide a benchmark and design principle for asymmetric imaging across platforms, with implications for defense, surveillance and secure communications.

## 1. Introduction

The ability to control the direction of propagation of light is fundamental to a wide variety of optical technologies ranging from imaging and spectroscopy to augmented and virtual reality. The past two decades have witnessed the surge of interest in devices capable of discriminating light fields based on their direction of propagation. Optical isolators[1] are the prototypical example: by definition, they use a nonreciprocal response to transmit light preferentially in one direction while suppressing backward propagation. Nonreciprocity has been realized using magneto-optic bias[2], spatiotemporal modulation[3-10], and nonlinear state dependence[11-21].

By contrast, a reciprocal multimode system can exhibit different total transmitted powers for inputs from opposite sides when those inputs are not time-reversed mode pairs. Reciprocal structures with structural chirality have exploited this freedom to produce large asymmetric-transmission contrasts[22-27], but their scattering matrices remain reciprocal, and they do not provide true isolation. This distinction between nonreciprocal isolation and reciprocal transmission asymmetry is central to the present work. Reciprocal systems are attractive precisely because they avoid the magnetic bias, active modulation, or nonlinear operating-point requirements of nonreciprocal devices, but this simplicity comes with a trade-off: any reciprocal system that transmits power in the forward direction must also transmit power in the reverse direction for the corresponding time-reversed modes, posing a fundamental challenge for asymmetric imaging.

Nonreciprocity and asymmetric transmission have recently been explored also in the context of imaging devices, for instance in the context of nonreciprocal analog image processing and of unidirectional imagers[28-31]. These devices form or process an image of an object when illuminated from one side, while performing a different operation, or fully blocking the transmission, when illuminated from the other side. In particular, recent works have demonstrated reciprocal multilayer diffractive structures consisting of subwavelength scatterers to achieve unidirectional imaging in the visible[28] and terahertz[29] wavelength ranges. Such imagers are typically designed by treating the optical response of each scatterer as a trainable parameter that is optimized for imaging asymmetry over a predefined set of inputs. In some ways generalizing the idea of reciprocal transmission asymmetry discussed above, these devices are optimized to enable imaging for a set of images when illuminated from one side, while blocking the same images for the opposite illumination. Since this approach involves modeling the propagation of a spatially varying source field derived from the input through the system, it implicitly assumes coherent illumination. While a recent study has proposed diffractive asymmetric imagers for partially spatially incoherent illumination,[32] the achieved asymmetry is substantially reduced when the correlation length falls below 1.5 times the operating wavelength. These factors limit the universal applicability of previously proposed diffractive multilayers. They also leave open a key question: are there reciprocity-based limits to maximum achievable unidirectional imaging, in the same way there are fundamental limits to asymmetric transmission for reciprocal systems?

Here, we introduce a theoretical formalism to retrieve bounds on imaging asymmetry for linear, passive reciprocal optics. For both coherent and incoherent excitations, we identify fundamental limits on asymmetric imaging for $M$ orthogonal object patterns drawn from the $N$ well-coupled object/image modes of the system. Perfect unidirectionality is possible for $M \leq N/2$; for $M > N/2$, the maximum average power asymmetry decreases as $(N - M)/M$, vanishing at full modal occupation ($M = N$). This limit is ruled by reciprocity alone, independent of the imaging mechanism. In fact, we propose simple refractive imaging systems based on common bulk optical elements that can achieve unidirectionality at the half-occupation efficiency limits. Next, we demonstrate a 1 cm diameter all-silicon metasurface doublet (meta-doublet) for asymmetric focusing at a wavelength of 4 µm.

Guided by our theory, we engineer the point spread function (PSF) of the doublet to resemble a focal spot in the forward direction and a random intensity pattern in the backward direction. We fabricate our meta-doublet using direct laser writing and experimentally validate its asymmetric imaging performance at a wavelength of 4 μm. Our results offer a systematic understanding of how asymmetric imaging can be achieved in reciprocal optical systems, both refractive- and metasurface-based, with broad applications beyond imaging, including fiber-optics communication, on-chip computing, defense and surveillance.

## 2. Results and discussion

We begin by envisioning a general asymmetric imaging scenario, followed by a discussion of the theoretical framework describing it. We use our theory to draw conclusions on the conditions under which a given imager can exhibit perfect asymmetry and use these derived conditions to design refractive unidirectional imagers. Finally, we use our formalism to design a compact, metasurface doublet in the mid-infrared wavelength range, and experimentally validate its asymmetric imaging performance.

### 2.1 Theoretical formalism of asymmetric imaging in reciprocal optical systems

The basic principle of a reciprocal unidirectional imager is described by the schematic in Fig. 1. For generality, we represent the imager as a scattering medium between two finite planes, labeled as 1 and 2. A unidirectional imager aims at producing the image of an object placed at plane 1 in plane 2 (forward direction), while prohibiting the formation of the image on plane 1 for the same object placed in plane 2 (backward direction). The object in plane 1 lies within aperture *A*, and we consider the same aperture in plane 2 to demarcate where images are formed and where objects are illuminated in the reverse direction. Reciprocity means that the transmission amplitude from an input mode to an output mode is unchanged when the source and receiver are interchanged and both modes are time reversed. Thus, if an object mode at plane 1 forms an image mode at plane 2, the time reverse of that image mode necessarily couples back to the time reverse of the object mode. Perfect asymmetry for the same object launched from plane 2 is therefore possible only if that reverse object mode is orthogonal to the reciprocal partner of the forward image. For a single object, one can enforce this orthogonality simply by shifting the image to a nonoverlapping spatial region. For many orthogonal objects, however, the number of available orthogonal image modes becomes limited, giving rise to the bounds derived below.

Any linear imaging system defines a transmission operator from the modes supported within aperture *A* at the object plane to those supported within the corresponding aperture *A* at the image plane. These mode spaces may include arbitrary vector electromagnetic fields, including polarization, so the formalism is not restricted to scalar fields[33]. We denote these operators by $\mathcal{F}$ and $\mathcal{B}$ in the forward and backward directions. For finite-sized scatterers, these compact operators admit arbitrarily accurate matrix representations[34], which we denote by $\mathbb{F}$ and $\mathbb{B}$, respectively. Each matrix has a singular value decomposition (SVD), which for $\mathbb{F}$ we write as

$$\mathbb{F} = U\Sigma V^{\dagger} = \sum_{i=1}^{N} \sigma_i u_i v_i^{\dagger} \,, \tag{1}$$

with $v_i$ being basis functions for objects, the $u_i$ being measurement basis functions, the singular values $\sigma_i$ representing the strength of the signal received in $u_i$ (in descending order), and $N$ the smaller of the number of rows and columns of $\mathbb{F}$. The singular values for propagation through finite electromagnetic scatterers must exhibit quasi-exponential fall-off, with a sharp transition from order-unity values to exponentially small values[35,36]. For the purposes of the bounds below, we define $N$ as the number of singular values exceeding a chosen threshold set by the required detection sensitivity, thereby setting the effective number of well-coupled modes. For a spatially incoherent system, the transmission matrix acts on nonnegative intensity vectors. Its right singular vectors are generally signed and therefore need not correspond to physically realizable object patterns. Consequently, unlike in the coherent case, the number of singular values above threshold cannot automatically be interpreted as the number of independently excitable channels. Nevertheless, the same bound holds: a direct argument on the intensity transmission matrix (Supplementary Note 1) yields Eq. (4) for nonnegative object patterns. Saturating the bound additionally requires an accessible set of orthogonal, nonnegative object patterns[37], which we assume in the ideal case considered below. Signed contrast modes could instead be implemented using controlled bias illumination and differential measurements, but that protocol is not assumed here.

Reciprocity manifests as transpose symmetry: the backward transmission matrix is the transpose of the forward transmission matrix, so that we can write

$$\mathbb{B} = \mathbb{F}^T = V^* \Sigma \mathrm{U}^T = \sum_{i=1}^{N} \sigma_i v_i^* u_i^T \,. \tag{2}$$

The most important feature of Eq. (2) is that the forward and backward transmission matrices share the same singular value distributions: for every well-coupled forward-going channel, there is an equivalent backward channel. Moreover, the right (transmitter) singular vectors in the backward direction are the conjugates (time-reversal partners) of the left (receiver) singular vectors in the forward direction, and similarly for the receiver singular vectors in the backward direction. The sharp decay of singular values implies that only image components in the space of well-coupled singular vectors can be sent with high fidelity and/or reconstructed accurately. Of the combinatorially many images that can be sent (in either direction), all are linear combinations of the right singular vectors. Thus, if $N$ singular values lie above the threshold, the well-coupled image space is $N$-dimensional and supports at most $N$ mutually orthogonal image modes. These singular modes need not be localized pixels. For a diffraction-limited imager, $N$ can nevertheless be estimated as the number of independent spatial degrees of freedom supported across the view, commonly quantified by the system's space-bandwidth product[38,39].

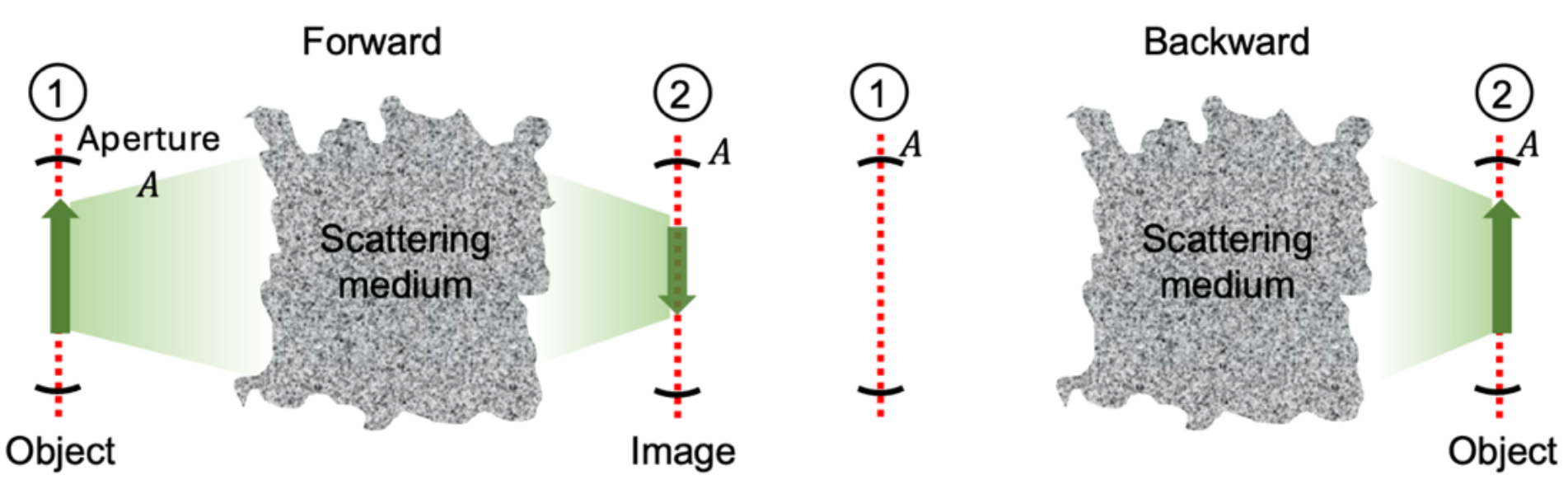

**Figure 1:** Schematic showing the operation of a reciprocal unidirectional imager.

Physical scrambling of a forward image may not be sufficient to prevent imaging in the reverse direction. In particular, if the transmission matrix (or the singular values and singular vectors of the well-coupled channels) is known, then it is feasible to decompose the measured intensities against the receiver-side singular vectors to recompose the image. In the presence of additive noise, the maximum-likelihood reconstruction of each object component is limited only by the signal-to-noise ratio of its channel: the Fisher information for the amplitude carried by singular vector $i$ scales as $\sigma_i^2/(\text{noise power})$, so the Cramér–Rao bound on estimating that component is set entirely by the received power $\sigma_i^2$, and ordinary least-squares reconstruction approaches that bound in the absence of prior information. A channel delivering order-unity power can therefore be inverted with high fidelity, whereas a channel delivering negligible power is irrecoverable, no matter how precisely the receiver knows the transmission matrix. Spatial scrambling thus provides no protection against a receiver that knows $\mathbb{B}$; the only quantity that fundamentally limits reconstruction is the received power.

We consider this general context in which both "friend" (detecting at plane 2) and "foe" (detecting at plane 1) know the transmission matrix to sufficiently high accuracy, possibly by training against image-object data. In this setting, the only way to achieve an asymmetry in imaging capability is to achieve a *power asymmetry*: to have objects from plane 1 produce order-unity power images at plane 2, but objects in plane 2 produce little to no power in plane 1. For a set of *M* objects for which unidirectional imaging is desired, the average power asymmetry is

$$\Delta P = \frac{1}{M}\sum_{i=1}^{M}\left(P_{f,i} - P_{b,i}\right), \tag{3}$$

where $P_{f,i}$ and $P_{b,i}$ are the received powers at planes 2 and 1, when image *i* is sent with unity power in the forward and backward directions, respectively. With unit-power inputs and passive optics ($\sigma_i \leq 1$), each $P_{f,i}$ and $P_{b,i}$ are in [0,1], so $\Delta P$ is in [-1,1], with $\Delta P = 1$ and $\Delta P = -1$ denoting perfect unidirectionality in the forward and backward directions, respectively. Throughout, we designate the forward direction as the desired one and therefore maximize $\Delta P$. We quantify asymmetry by this power difference rather than the forward/backward power ratio conventionally used for isolators because, by the argument above, it is the received power that bounds an informed adversary's reconstruction capability.

By linearity, we can analyze *M* orthogonal object patterns, and our efficiency ratios apply equally to arbitrary linear combinations. If *M* equals 1, perfect asymmetry is possible. A single-pixel object at (say) the center of plane 1 can scatter away from the center of plane 2; then, there is no reciprocity constraint on an object placed at the center of plane 2, which can then be scattered elsewhere, outside the aperture *A* of plane 1. In the other limit, consider the case *M* = *N*: the object set spans the entire well-coupled modal space, completely preventing unidirectionality, as depicted in Fig. 2. The total power sent across all channels in the forward direction must exactly equal the total power sent across all channels in the reverse direction, due to the identical singular-value distributions. Hence, in the following we explore the extent to which unidirectionality (or asymmetry) can be achieved for intermediate values of *M* as constrained by reciprocity.

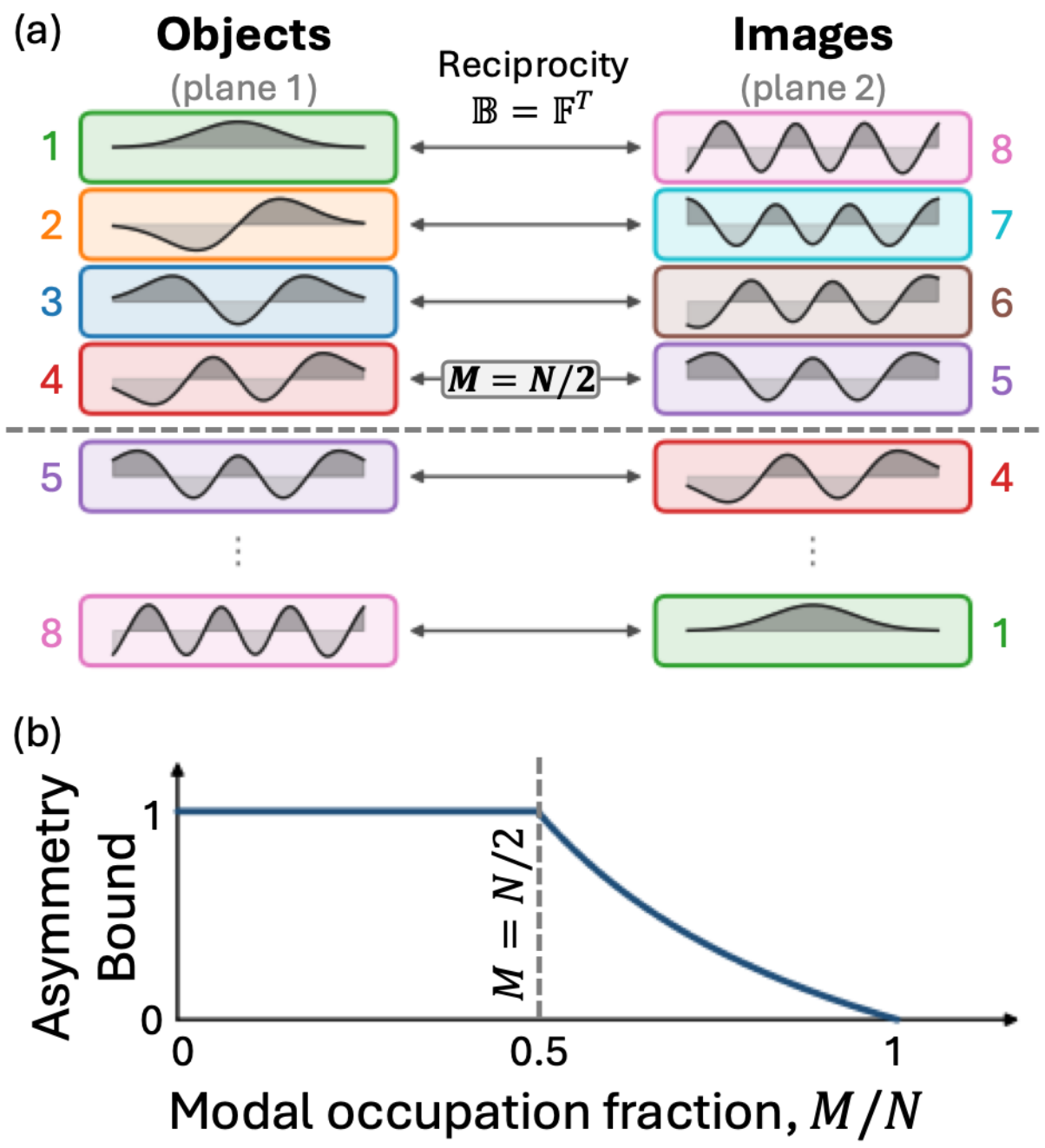

**Figure 2:** Reciprocity bounds the attainable imaging asymmetry. (a) Picture underlying Eq. (4), shown for N=8 degrees of freedom drawn as orthogonal modes rather than pixels. The forward transmission matrix couples object modes at plane 1 (left) to image modes at plane 2 (right); reciprocity makes the reverse map its transpose. Matched colors map symmetric partners. An object set can achieve perfect asymmetry only while it stays below half occupation; beyond $M = N/2$ (dashed line) further objects must reuse already-occupied degrees of freedom, which then transmit symmetrically. (b) Resulting bounds on the average power asymmetry [Eq. (4)] versus the modal occupation fraction $M/N$ – the fraction of the degrees of freedom occupied by objects. The bounds decay as $(N - M)/M$ to zero at $M = N$.

We find that for any values of *M* and *N*, for coherent or incoherent sources, average power asymmetry is bounded as (Fig. 2(b))

$$\Delta\mathrm{P} \leq \begin{cases} 1 & M < \dfrac{N}{2} \\ \dfrac{N-M}{M} & M \geq \dfrac{N}{2}. \end{cases} \tag{4}$$

Here we prove this bound for the coherent case; we prove the incoherent case in Supplementary Note 1. In the coherent case, an input object can be encapsulated in an *N x 1* vector $\boldsymbol{x}$, the output measurement $\boldsymbol{y}$, and each can be normalized such that their transmitted/received powers are $\boldsymbol{x}^\dagger\boldsymbol{x}$ and $\boldsymbol{y}^\dagger\boldsymbol{y}$, respectively. Object $\boldsymbol{x}_i$ transmitted in the forward/backward directions produces $\boldsymbol{y}_\mathbb{F} = \mathbb{F}\boldsymbol{x}_i$ and $\boldsymbol{y}_\mathbb{B} = \mathbb{B}\boldsymbol{x}_i$, respectively, such that the power asymmetry is $\boldsymbol{y}_\mathbb{F}^\dagger\boldsymbol{y}_\mathbb{F} - \boldsymbol{y}_\mathbb{B}^\dagger\boldsymbol{y}_\mathbb{B}$. Averaged over all *M* objects and inserting the SVDs for $\mathbb{F}$ and $\mathbb{B}$, the power asymmetry is

$$\Delta P = \frac{1}{M}\sum_i^M \boldsymbol{x}_i^\dagger (V\Sigma^2V^\dagger - U^*\Sigma^2U^T)\boldsymbol{x}_i = \frac{1}{M}\mathrm{Tr}\left[(V\Sigma^2V^\dagger - U^*\Sigma^2U^T)\sum_i \boldsymbol{x}_i\boldsymbol{x}_i^\dagger\right], \tag{5}$$

where "Tr" denotes matrix trace. If *M* = *N*, then the outer product term on the mutually orthogonal $\boldsymbol{x}_i$ vectors must be the *N x N* identity matrix, each term simplifies to $\mathrm{Tr}(\Sigma^2)$, since $U$ and $V$ are unitary, and the average power asymmetry equals zero, as predicted in the discussion above. For arbitrary *M*, the outer product summation corresponds to a rank-*M* matrix with all eigenvalues equal to unity. Each term is of the form $\mathrm{Tr}(WAW^\dagger B)$ for unitary $W$. By von Neumann's trace inequality[40,41], such terms satisfy bounds, $\lambda^\downarrow(A)\lambda^\uparrow(B) \leq \mathrm{Tr}(WAW^\dagger B) \leq \lambda^\downarrow(A)\lambda^\downarrow(B)$, where the superscripts denote descending ($\lambda^\downarrow$) or ascending ($\lambda^\uparrow$) ordering of the respective matrix eigenvalues. The essence of these bounds is that the product of one matrix with the unitary transformation of another can range from maximally "misaligned" eigenvalues (descending eigenvalues of one aligned with ascending of the other) to maximally aligned (descending together). Denoting $X = \sum_i \boldsymbol{x}_i\boldsymbol{x}_i^\dagger$, we can bound the first (positive) term of Eq. (5) with the trace upper bound, and the second (negative) term with the lower bound, producing

$$\Delta P \leq \frac{1}{M}\left[\lambda^\downarrow(\Sigma^2)\lambda^\downarrow(X) - \lambda^\downarrow(\Sigma^2)\lambda^\uparrow(X)\right] = \frac{1}{M}\left[\sum_i^M \sigma_i^2 - \sum_{N-M+1}^N \sigma_i^2\right], \tag{6}$$

which is simply the difference between the average of the first *M* singular values $\sigma_i^2$ and the last *M* squared singular values, of which there are *N*. This bound has a similar incoherent analog, without the same quadratic forms appearing, but leading to the same bound, as derived in Supplementary Note 1. For singular values bounded above by 1 (unity power normalization and passivity), the bound of Eq. (6) simplifies exactly to the bound of Eq. (4). This bound is a fundamental limit that cannot be surpassed by any passive optical imaging system, ray-based or wave-based, if it is linear and reciprocal.

The bound of Eq. (4) can be intuitively understood through a generalization of the classical notion of étendue. Just as the product of solid angle with spatial area, $d\Omega dA$, forms a differential étendue in geometric optics, each communication channel forms a discrete étendue element in wave optics[42]. The reciprocity relation $\mathbb{B} = \mathbb{F}^T$ enforces a wave-optical étendue equality: the forward and reverse transmission matrices have identical singular values and therefore the same number $N$ of well-coupled modes. This reciprocity-based equality should not be confused with étendue conservation, which may fail in the presence of finite apertures or loss. The derivation does not assume losslessness: absorption and scattering outside the designated apertures are both represented by reduced singular values, and passivity requires only $\sigma_i \leq 1$. Equivalently, every well-coupled forward mode has a reciprocal reverse partner. For $M \leq N/2$, the object modes can be mapped to image modes that do not participate in the reverse object excitation, allowing perfect asymmetry. For $M > N/2$, the forward image subspace must overlap the reverse object subspace, so the maximum asymmetry decreases and vanishes at full modal occupation ($M = N$).

**2.2 Unidirectional imaging using refractive optics**

Here, we present two examples of unidirectional imagers, each constructed from a biconvex lens and a stop, that attain or approach the half-occupation limit of the previous section. The first example is shown in Fig. 3(a) and consists of a single biconvex lens and a stop placed to its right, extending infinitely above the optic axis in a direction perpendicular to it. Let us consider imaging at the planes labeled as 1 and 2, with their distances to the lens related by the lens equation. For an object placed above the optic axis on plane 1, the lens produces a real and inverted image on plane 2. We regard this as the forward direction and denote it as 1→2. In the backward direction 2→1, the same object placed on plane 2 is completely blocked by the stop and is unable to produce an image on plane 1. Therefore, the system achieves perfect imaging asymmetry for the given set of planes 1 and 2.

We note that the imager is asymmetric for all choices of plane 1 to the left of the lens, provided 1→2 is considered as the forward direction, and the object is placed above the optic axis. On the other hand, the imager is also asymmetric for all choices of plane 2 provided 2→1 is considered as the forward direction, and the object is placed below the optic axis. These two cases align with the conclusions of the asymmetry analysis above. Considering each basis vector on a given plane as a single-pixel object, the imager presented in Fig. 3(a) achieves perfect asymmetry only for objects occupying at most half of the available degrees of freedom – here, the spatial pixels (modes) of plane 1 or 2. On the other hand, objects occupying more than half of those degrees of freedom on a given plane are not fully imaged by the imager, thus causing it to have partial asymmetry.

Figure 3(b) presents a second mechanism for asymmetric imaging, magnification, which approaches perfect asymmetry through the same area–angle tradeoff that underlies our theory. The system is a single thin lens of spatial magnification $|m| \ll 1$ with a stop at the image plane. In the forward direction (1→2), the lens maps a spatially extended object that spans a large area but a small range of angles (the wide, low strip of phase space in the plane-1 panel of Fig. 3(b)) onto a demagnified image that occupies a small area but a large range of angles (the narrow, tall strip in the plane-2 panel). Because étendue is conserved, demagnifying the image by $|m|$ in space magnifies its angular content by $1/|m|$ — the compact forward image passes cleanly through the stop. In the backward direction (2→1), an object placed at plane 2 spans the full aperture at small angles (the wide, low strip in the plane-2 panel).

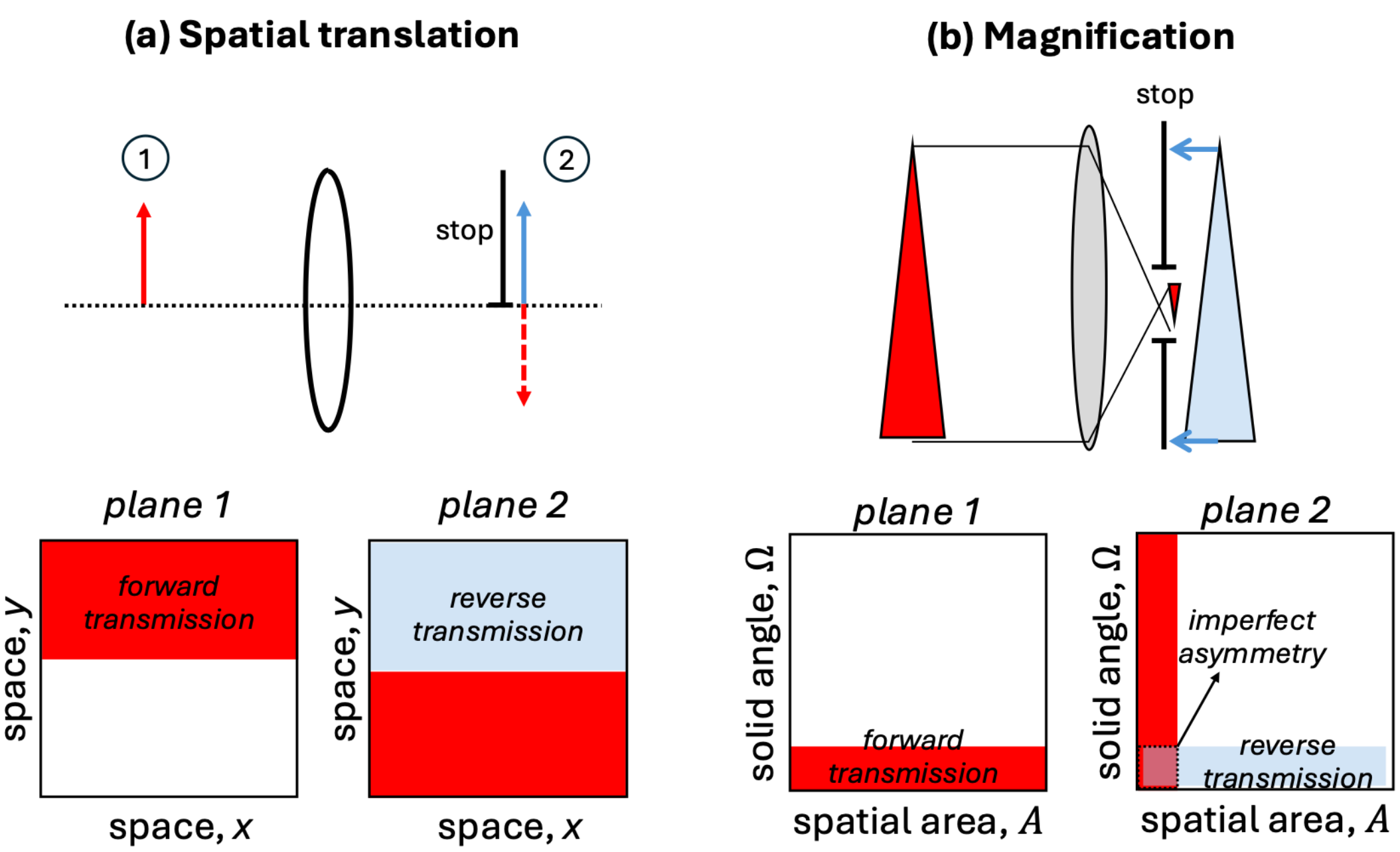

**Figure 3:** Two refractive mechanisms for unidirectional imaging. Top panels: ray diagrams. Bottom panels: occupied degrees of freedom at planes 1 and 2 (forward red, reverse blue). (a) **Spatial translation**. A biconvex lens with a stop images the upper half of plane 1 onto the lower half of plane 2; objects at plane 2 are blocked and send no power back. In transverse position space (lower panels), forward images and reverse objects occupy disjoint DoFs — the largest fraction of phase space (one half) compatible with unity efficiency — so the asymmetry is perfect. (b) **Magnification**. A thin lens of magnification $|m| \ll 1$ with a stop at the image plane maps a large area with small angular range onto a small area with large angular range (lower panels, in area–angle phase space). Reverse objects now overlap the forward images in the small-area, small-angle corner, with the (small) fraction $|m|$ of the reverse-directed power reaching plane 1, giving a near-ideal asymmetry $\Delta P = 1 - |m|$.

This reverse object overlaps the forward-going channels only in the region that is simultaneously small in area and small in angle (the shaded corner of the plane-2 panel), so only a fraction $|m|$ of its power couples into channels that can be reconstructed at plane 1. The imager therefore achieves a power asymmetry $\Delta P = 1 - |m|$, approaching unity as $|m| \to 0$.

This places the magnification imager just inside the half-occupation regime of Eq. (4): it approaches the unity-asymmetry limit, with the residual reverse coupling set by the unavoidable phase-space overlap rather than by any reciprocity violation. In contrast to the translation imager of Fig. 3(a), which attains perfect asymmetry exactly over half the aperture, the magnification imager offers a continuous knob that trades a small, controllable loss of asymmetry for the ability to image at finite magnification.

A useful feature of our asymmetry theory is the fact that it is agnostic to the set of basis vectors for the two planes. Consequently, one may choose to look at the imaging problem in the Fourier space instead of the real space. This is because the light propagating from a given object towards the imager can always be decomposed into a set of plane waves with distinct wavevectors. It can then be said that a given system can achieve perfect asymmetry only for objects that can be described using at most half of the available degrees of freedom – here, the plane-wave modes (Fourier wavevectors) of plane 1.

### 2.3 Meta-doublet for asymmetric imaging

Metasurfaces offer both compactness and subwavelength spatial control of optical phase. In multi-surface systems, propagation between patterned surfaces enables richer transformations between input and output spatial modes than the simple lens-and-stop systems considered above. We exploit this mode-control freedom to design a metasurface doublet (meta-doublet) for asymmetric focusing of normally incident plane waves (Fig. 4). The imager consists of two axially separated metasurfaces (meta-optics) that focus a plane wave at the wavelength of 4 µm normally incident from the left to a distance of 1 cm to their right (left panel of Fig. 4). Conversely, for a normally incident plane wave from the right, the system produces a random pattern at a distance of 1 cm to their left (right panel of Fig. 4). The two patterned surfaces and the propagation between them provide the degrees of freedom needed to tailor the forward and backward modal transformations differently, subject to reciprocity.

Before discussing the specific design of the meta-doublet, it is useful to view it in the context of the asymmetry theory outlined above. For the parallel wavevector $\vec{k}_0 = \vec{0}$ (incident from plane 1), the doublet produces a focal spot on plane 2. We note that a focal spot at the center of the image plane is the Fourier transform of a normally incident plane wave. Hence, in the forward direction the doublet maps the on-axis input mode (the $\vec{k}_0 = \vec{0}$ plane wave) to the on-axis output mode (the central focal spot) — the M = 1 single-mode case of Sec. 2.1, for which perfect asymmetry is permitted. In the backward direction, a normally incident plane wave scatters to a set of higher magnitude wavevectors, since the doublet produces a random pattern for a normally-incident plane wave. The doublet can thus be considered as an imperfect asymmetric imager since a perfect one would need to completely block the light from reaching plane 1 in the backward direction. In Supplementary Note 5, we present the design of a doublet that achieves near-perfect unidirectional imaging of normally incident plane waves.

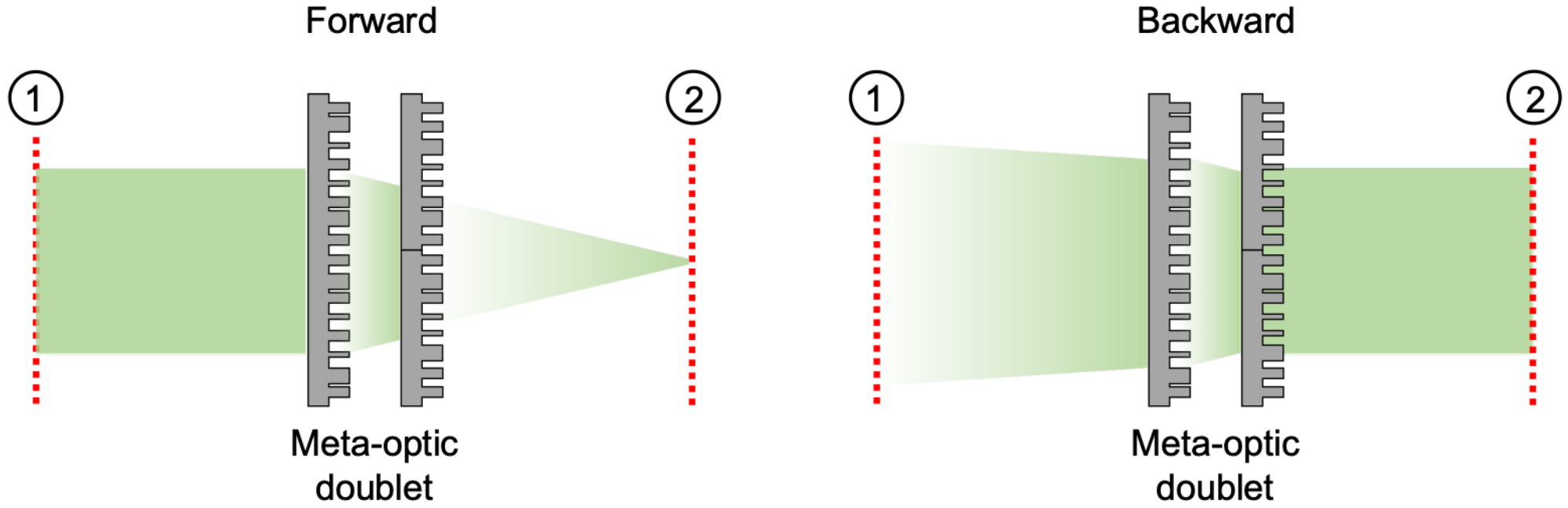

**Figure 4:** Schematic showing the working principle of an asymmetric imaging meta-doublet.

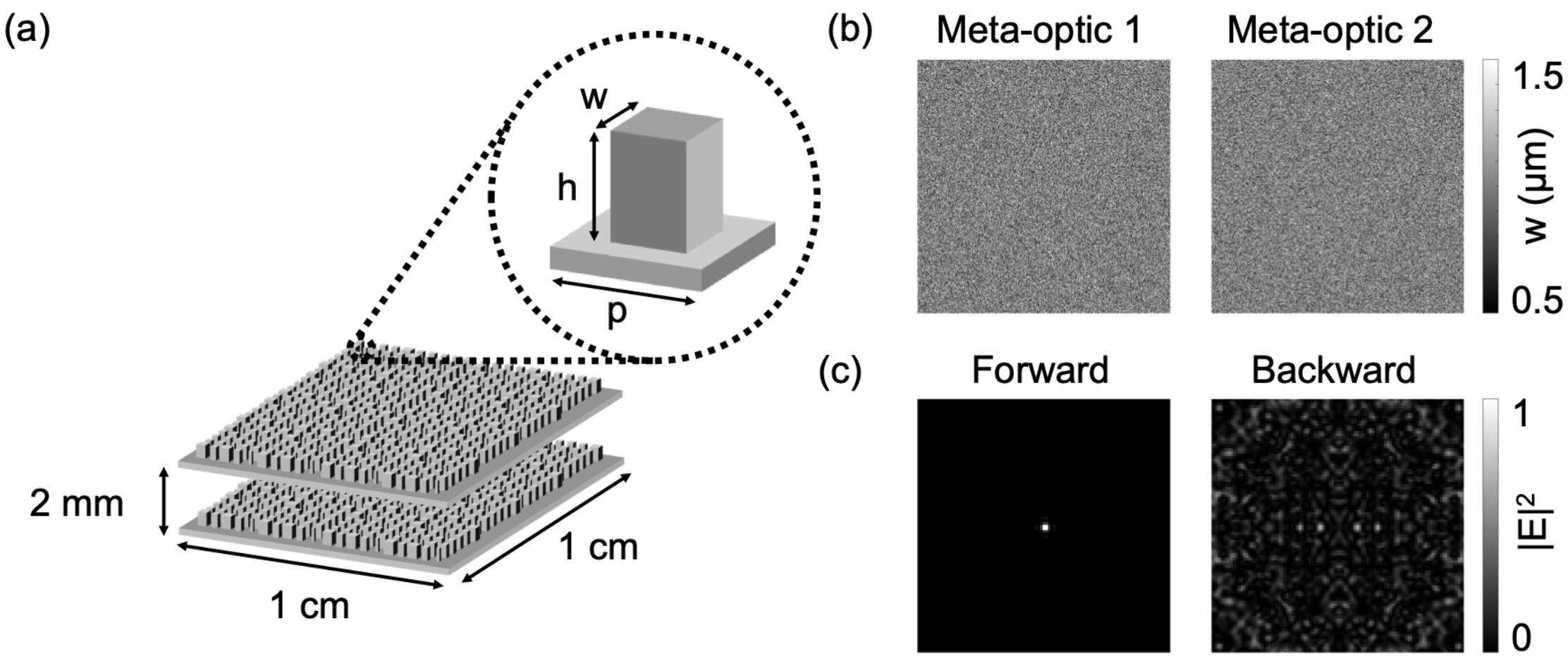

**Figure 5:** (a) Schematic of the designed all-silicon metasurface doublet with a single constituent nanopillar shown in the inset. (b) Optimized feature maps of the meta-optics used in the doublet. (c) Simulated point spread functions (central 0.4 x 0.4 $mm^2$ region) of the doublet in the forward and backward directions at a wavelength of 4 μm. The PSFs are normalized to their own maximum.

The meta-optics forming the doublet have a diameter of 1 cm and are separated by a distance of 2 mm (Fig. 5(a)). We implement the metasurfaces using all-silicon square scatterers with a height *h* of 3 μm and a center-to-center separation of 2 μm (inset of Fig. 5(a)). The widths of the scatterers vary between 0.5 and 1.5 μm as determined by fabrication constraints. The two meta-optics are designed using a gradient descent-based optimization with the scatterer widths as the free parameters. For computational efficiency, we impose a four-fold symmetry on the structure of both meta-optics. The imaging performance of the doublet in the two directions is quantified by its Strehl ratio, defined here as the integral of the modulation transfer function (MTF) normalized to that of the corresponding diffraction-limited optic; a value of 1 therefore represents diffraction-limited performance. The optimization leverages a differentiable mapping between the scatterer widths and the point spread functions (PSFs) of the doublet, from which the Strehl ratios are calculated. The loss function quantifies the imaging asymmetry of the doublet in the forward and backward directions and is given by the ratio of their corresponding Strehl ratios. Supplementary Note 2 provides further details of the meta-doublet design process.

The optimized feature maps of the two meta-optics are shown in Fig. 5(b). The distribution of pillar widths appears to be completely random since the four-fold symmetry planes are not visible at this scale. The PSFs of the doublet in the forward and backward directions calculated using the angular spectrum method are shown in Fig. 5(c). For ease of visualization, both PSFs are normalized to their own maximum and only the central 0.4 x 0.4 $mm^2$ regions are displayed. As designed, the doublet produces a clear focal spot in the forward direction and a random intensity pattern in the backward direction. The Strehl ratios of the doublet in the forward and backward directions are 0.13 and 8.6 x $10^{-4}$. In Supplementary Note 6, we show that the asymmetric focusing performance of the doublet is a consequence of the weakly modulating nature of the constituent meta-optics and their complementary focal lengths.

The constituent meta-optics are fabricated using direct laser writing on a single chip and later separated via dicing. Figure 6(a) presents an image of the initial fabricated chip taken with a phone camera. The optical microscope image of one of the meta-optics is shown in Fig. 6(b) and the corresponding scanning electron microscope (SEM) image is shown in Fig. 6(c). Figure 6(d) presents the experimentally measured point spread functions of the doublet in the forward and backward directions, normalized to their own maximum. Supplementary Notes 3 and 4 provide details of the meta-doublet fabrication and characterization. As designed, the doublet produces a clear focal spot in the forward direction and a random PSF in the backward direction. In Fig. 6(e), we plot the intensity at the center of the PSF in the forward and backward directions as a function of the distance *z* from the trailing optic of the doublet. The forward intensity peaks at z = 9.5 mm, within 5% of the designed focal distance of 10 mm. A Gaussian fit gives an axial FWHM of 0.66 mm for the focusing-response peak; this quantity characterizes the axial extent of the asymmetric-focusing regime rather than the transverse focal-spot width. The simulated forward Strehl ratio of 0.13 indicates that the design is not diffraction limited. In Supplementary Note 7, we present results of imaging simulations conducted using the numerically calculated PSFs of the doublet, providing further evidence for its asymmetric imaging performance.

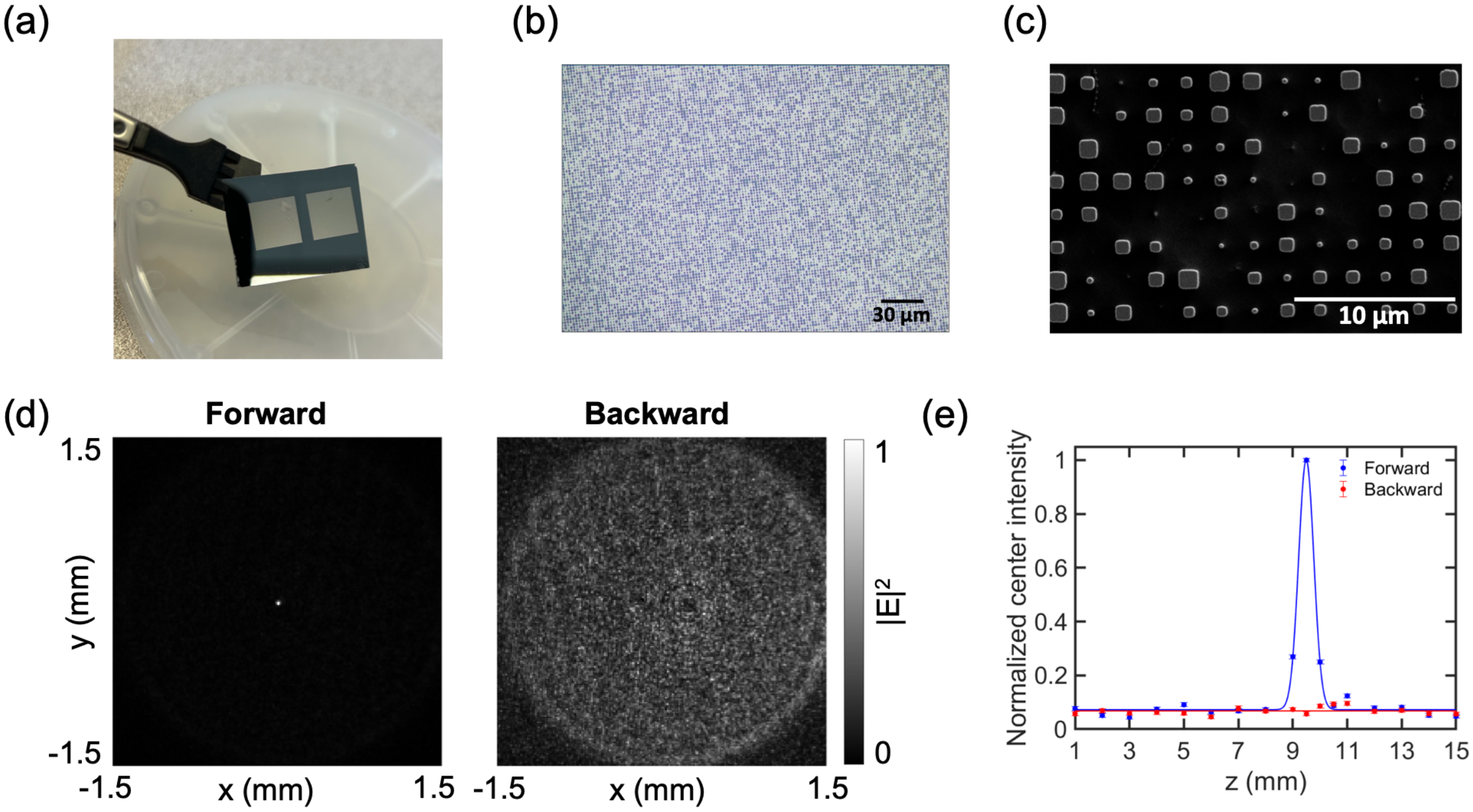

**Figure 6:** (a) Image of the chip with the fabricated meta-doublet captured with a phone camera. (b) Optical microscope, and (c) scanning electron microscope image of one of the optics in the doublet. (d) Measured PSFs of the doublet in the two directions at a wavelength of 4 µm and at a distance of 9.5 mm. The PSFs are normalized to their own maximum. (e) Intensity at the center of the PSF in the forward and backward directions as a function of the distance from the doublet *z*. The forward direction intensities are fitted to a Gaussian function (solid blue curve) while a solid red line at the mean backward direction intensity level is plotted as a guide to the eye.

## 3. Conclusion

In this work we developed a general theory for asymmetric imaging in linear, passive, reciprocal optical systems. Representing any such imager through the singular-value decomposition of its forward transmission matrix and invoking reciprocity ($\mathbb{B} = \mathbb{F}^T$), we derived a fundamental bound [Eq. (4)] on the average power asymmetry achievable for any M objects drawn from a system with N degrees of freedom, valid for coherent and incoherent illumination alike. The bound shows that perfect asymmetry is attainable only at or below half modal occupation ($M \leq N/2$) and decays as (N−M)/M beyond it — a limit set purely by reciprocity, independent of physical mechanism, that no passive reciprocal imager can surpass. We then illustrated the theory with two complementary realizations: refractive imagers of lenses and stops that lay bare the underlying area–angle (étendue) tradeoff, and a large-area, all-silicon metasurface doublet that we designed, fabricated and experimentally validated in the mid-infrared.

Our all-silicon metasurface doublet (fabricated by direct laser writing) is capable of asymmetric focusing in the mid-infrared, achieved by engineering the point spread function of the doublet to resemble a focal spot in the forward direction and a random intensity pattern in the reverse direction, at a wavelength of 4 µm. In principle, it is possible to design a meta-doublet that achieves asymmetric imaging for many distances and a broad range of wavelengths. However, this would require optimizing the focusing asymmetry for a broad range of incident angles and wavelengths, thus becoming computationally expensive for large area optics. Additionally, the optical response of the doublet is sensitive to the spacing between the meta-optics, which can potentially be overcome by maximizing its depth of focus in addition to its imaging asymmetry during the design process.

Looking forward, the theoretical framework offers several applications and extensions. Because the bound follows from linearity and reciprocity, it applies to any linear, passive, reciprocal asymmetric imager—refractive, diffractive, or metasurface-based—and provides both a benchmark and a design principle: asymmetry is maximized by operating below half modal occupation and concentrating object information in the most strongly coupled channels. The framework places asymmetric imaging within the established context of optical degrees of freedom, space–bandwidth product, and étendue, recasting unidirectionality from a device-specific effect into a resource constrained by reciprocity. Breaking reciprocity[31] removes the transpose constraint $\mathbb{B} = \mathbb{F}^T$, thereby permitting more general asymmetric transformations that in principle can surpass Eq. (4). Other directions include using the bound as an objective in end-to-end imager design and extending the analysis to broadband and partially coherent operation. We anticipate these limits will guide the design of next-generation asymmetric imaging systems across the electromagnetic spectrum.

## Acknowledgements

This work was supported by the DARPA Coded Visibility STTR program (Contract No: HR001123C0034).

## Author contributions

A.M. and O.M. conceived the project and supervised the work. R.A. and M.Z. designed and optimized the meta-doublet and performed the imaging simulations. O.M. developed the reciprocity bound, with input from R.P. and A.A. A.W.-S., R.J., and V.Y. fabricated the meta-optics and performed the optical characterization. All authors discussed the results and contributed to writing the manuscript.

**Competing interests**

A.M. declares a financial interest in Tunoptix, Inc., a company developing software-defined meta-optics. All other authors declare no financial or non-financial competing interests.

**Data availability**

The data that support the findings of this study, including the measured and simulated point spread functions and the optimized meta-optic width maps, are available from the corresponding authors upon reasonable request.

**Code availability**

The gradient-descent design code, which builds on the open-source packages S4 and waveprop, is available from the corresponding authors upon reasonable request.

# Supplementary Information for "Asymmetric imaging: Reciprocity bounds and manifestations"

Romil Audhkhasi[1], Anna Wirth-Singh[1], Rose Johnson[1], Maksym Zhelyeznyakov[1], Vladimir Yarmolik[2], Rafael Piestun[3,4], Andrea Alù[5,6], Arka Majumdar[1,2,*] and Owen Miller[7,8,*]
[1]Department of Electrical and Computer Engineering, University of Washington, Seattle, Washington 98195, United States
[2]Department of Physics, University of Washington, Seattle, Washington 98195, United States
[3]Department of Electrical, Computer and Energy Engineering, University of Colorado, Boulder, Colorado 80309, United States
[4]Department of Physics, University of Colorado, Boulder, Colorado 80309, United States
[5]Photonics Initiative, Advanced Science Research Center, City University of New York, New York, NY 10031, United States
[6]Physics Program, Graduate Center, City University of New York, New York, NY 10016, United States
[7]Department of Applied Physics, Yale University, New Haven, Connecticut 06520, United States
[8]Department of Electrical and Computer Engineering, Yale University, New Haven, Connecticut 06520, United States
[*]Corresponding authors: arka@uw.edu, owen.miller@yale.edu

## Supplementary Note 1. Asymmetry bound for spatially incoherent imaging

For spatially incoherent illumination, objects and images are described by nonnegative intensity distributions rather than by complex field amplitudes. We discretize the aperture *A* on each plane into *N* resolution cells (pixels) and describe an object by the *N* x 1 vector $\boldsymbol{x}$ of pixel intensities, normalized to unit power, $\mathbf{1}^T\boldsymbol{x} = 1$, where $\mathbf{1}$ denotes the all-ones vector. In this basis the forward transmission matrix $\mathbb{F}$ is the *N* x *N* matrix of nonnegative intensity transfer coefficients: $\mathbb{F}_{jk}$ is the power received in pixel *j* of plane 2 when unit power is launched from pixel *k* of plane 1. Each intensity coefficient is the squared magnitude of the corresponding field coefficient, so the transpose symmetry of the field transmission matrices [Eq. (2) of the main text] carries over to the intensity transmission matrices, $\mathbb{B} = \mathbb{F}^T$. The powers received when the object $\boldsymbol{x}$ is launched in the forward and backward directions are therefore

$$P_f = \mathbf{1}^T\mathbb{F}\boldsymbol{x}, \qquad P_b = \mathbf{1}^T\mathbb{B}\boldsymbol{x} = \mathbf{1}^T\mathbb{F}^T\boldsymbol{x} = \boldsymbol{x}^T\mathbb{F}\mathbf{1}. \tag{S1}$$

In contrast to the coherent case, no quadratic forms appear: the received powers are linear in the object vector. It is convenient to define, for each pixel *k*, the forward transmittance $t_k = (\mathbf{1}^T\mathbb{F})_k = \sum_j \mathbb{F}_{jk}$, the fraction of unit power launched from pixel *k* of plane 1 that reaches the aperture *A* of plane 2, and the backward transmittance $t'_k = (\mathbb{F}\mathbf{1})_k = \sum_j \mathbb{F}_{kj}$, the same quantity for pixel *k* of plane 2. Passivity requires $0 \le t_k \le 1$ and $0 \le t'_k \le 1$. The two sets of transmittances have equal sums,

$$\sum_{k=1}^{N} t_k = \sum_{k=1}^{N} t'_k = \mathbf{1}^T\mathbb{F}\mathbf{1}, \tag{S2}$$

which is the incoherent counterpart of the shared singular-value spectrum of Eq. (2): a uniformly illuminated aperture transmits the same total power in the two directions.

For a set of *M* objects $\boldsymbol{x}_i$, Eq. (3) of the main text becomes, with Eq. (S1),

$$\Delta P = \frac{1}{M}\sum_{i=1}^{M}(\mathbf{1}^T\mathbb{F}\boldsymbol{x}_i - \boldsymbol{x}_i^T\mathbb{F}\mathbf{1}) = \frac{1}{M}\sum_{k=1}^{N} s_k\,(t_k - t_k'), \qquad \boldsymbol{s} = \sum_{i=1}^{M}\boldsymbol{x}_i\,. \tag{S3}$$

The average power asymmetry thus depends on the object set only through the summed intensity pattern $\boldsymbol{s}$. Mutually orthogonal nonnegative patterns have disjoint supports, so each pixel belongs to at most one object, and $0 \le s_k \le 1$ with $\sum_k s_k = M$. Maximizing Eq. (S3) over all such $\boldsymbol{s}$ places unit weight on the *M* pixels with the largest values of $d_k = t_k - t_k'$:

$$\Delta P \le \frac{1}{M}\sum_{k=1}^{M} d_k^{\downarrow}\,, \tag{S4}$$

where $d_k^{\downarrow}$ denotes the $d_k$ arranged in descending order. We can then use two properties of the $d_k$ values. First, each $d_k \le 1$, so $\Delta P \le 1$. Second, by Eq. (S2) the $d_k$ sum to zero, so the sum of the *M* largest equals minus the sum of the *N* − *M* smallest, each of which is at least −1:

$$\sum_{k=1}^{M} d_k^{\downarrow} = -\sum_{k=M+1}^{N} d_k^{\downarrow} \le N - M. \tag{S5}$$

Combining the two properties gives $\Delta P \le \min\{1,\ (N-M)/M\}$, which is Eq. (4) of the main text. As in the coherent case, *N* counts the well-coupled pixels: pixels with $t_k = t_k' = 0$ contribute to neither sum and may be dropped. The derivation works directly with nonnegative intensities and does not require relaxing the nonnegativity constraint; it complements the coherent derivation of the main text, in which the objects are complex field patterns and the singular value decomposition of Eq. (1) is used.

Equation (S5) also identifies the configurations that saturate the bound. For $M \le N/2$, equality requires $t_k = 1$ and $t_k' = 0$ on the *M* object pixels, i.e., each object pixel is transmitted with unit efficiency in the forward direction and completely blocked in the backward direction, which by Eq. (S2) requires at least *M* other pixels with $d_k < 0$, i.e., with net backward transmission. For $M > N/2$, the maximum $(N-M)/M$ requires *N* − *M* pixels with $d_k = 1$, *N* − *M* pixels with $d_k = -1$, and the remaining 2*M* − *N* pixels transmitting symmetrically, $d_k = 0$. The lens-and-stop imager of Fig. 3(a) of the main text realizes the *M* = *N*/2 case: pixels in the upper half of the aperture have $(t_k,\ t_k') = (1,\ 0)$ and pixels in the lower half have $(t_k,\ t_k') = (0,\ 1)$.

## Supplementary Note 2. Details of the meta-doublet design and optimization

The meta-doublets presented in this work are designed using gradient-descent-based optimization. This requires the creation of a differentiable mapping between the structures of the doublets and their far-field responses, i.e., point spread functions (PSFs). In the context of this work, the PSF of a doublet in a given direction is the spatially varying intensity produced by it at the corresponding focal plane when illuminated by a normally incident plane wave. Since the meta-optics forming the doublets are made up of nanopillars with a fixed height and center-to-center separation, the only optimizable structural parameters are their widths.

We begin the optimization setup by creating a mapping between the widths of the scatterers and their near-field optical response. Supplementary Fig. 1 presents the transmission amplitudes and phases of silicon square scatterers having a height of 3 µm, center-to-center spacing of 2 µm and widths varying between 0.5 and 1.5 µm at a wavelength of 4 µm. These calculations are performed using the open-source rigorous coupled-wave analysis solver S4. We observe that the nanopillars are able to achieve the full 0 to 2π phase coverage for this range of pillar widths. Although the transmission amplitudes are low for nanopillar widths ranging from approximately 1.1 to 1.37 µm, we note that a vast majority of the nanopillars in the final designed meta-optics have high transmission amplitudes. To create a differentiable mapping, we fit the calculated phases and transmission amplitudes of the nanopillars to their widths using piecewise cubic polynomial functions. For a given distribution of nanopillar widths, and hence given near-field responses of the meta-optics forming a doublet, its PSF can be calculated using the angular spectrum method. Here, we use the differentiable angular spectrum propagator available from the open-source package waveprop for calculating the PSFs. This, along with the fitted polynomial functions for the nanopillar near-field responses, constitutes a fully differentiable wave-propagation pipeline for the doublets.

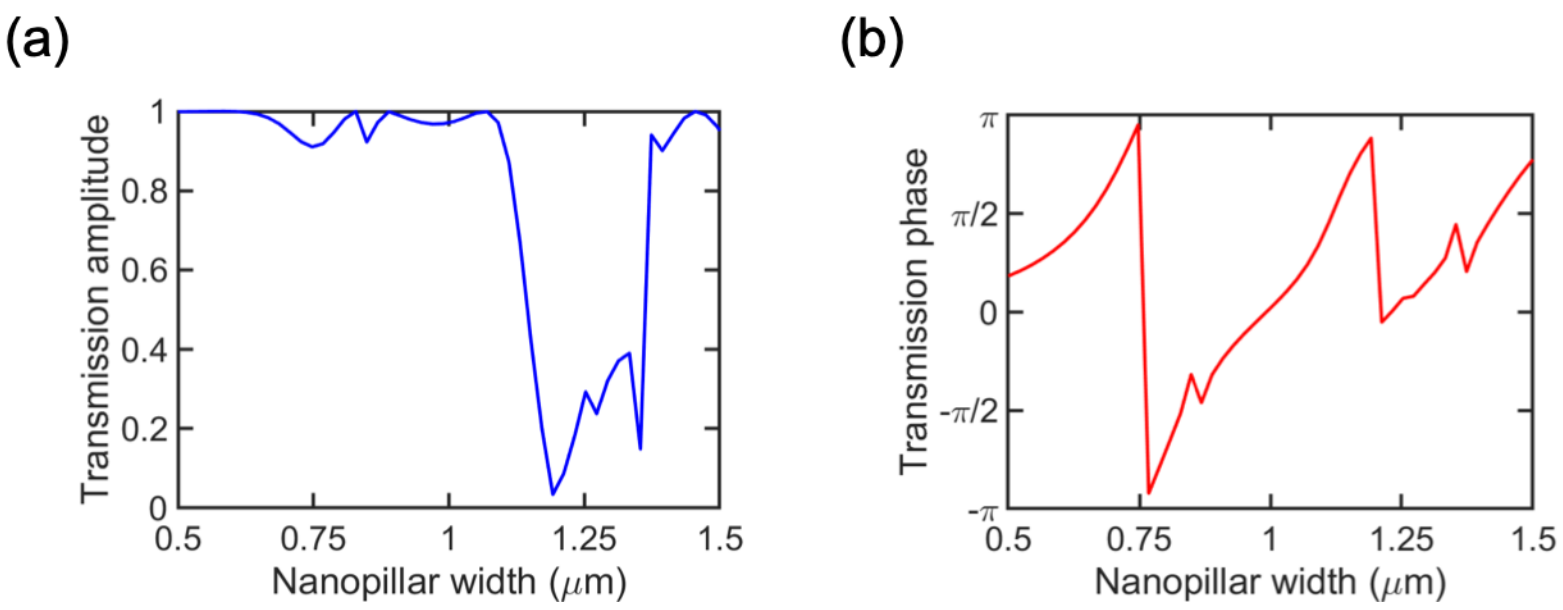

**Supplementary Figure 1:** (a) Transmission amplitudes, and (b) phases of the all-silicon square scatterers used for constructing the meta-optics in this work.

The final step of the optimization setup is the definition of a loss function. The loss function for the asymmetric imaging doublet is chosen to maximize its imaging performance in the forward direction relative to that in the backward direction. We quantify the imaging performance of the doublet using its Strehl ratio (SR), which is a number between 0 and 1, with 1 representing diffraction-limited imaging at a given wavelength. As in the main text, the Strehl ratio is defined as the integral of the modulation transfer function (MTF) normalized to that of the corresponding diffraction-limited optic:

$$SR = \frac{\iint dk_x dk_y \; MTF_{optic}\left(k_x, k_y\right)}{\iint dk_x dk_y \; MTF_{diff}\left(k_x, k_y\right)}. \tag{S6}$$

Here $MTF_{optic}$ is the modulation transfer function of the optic, given by the magnitude of the Fourier transform of its PSF, while $MTF_{diff}$ is the modulation transfer function of a diffraction-limited optic with the same aperture. The loss function for the asymmetric imaging doublet is defined as the ratio of the forward and backward SRs multiplied by the intensity at the center of the forward PSF. This ensures that the optimization favors a design that produces a focal spot in the forward direction and a random intensity pattern in the backward direction. For the unidirectional imaging doublet presented in Supplementary Note 5, the loss function is instead defined as the ratio of the total focal-plane powers in the two directions multiplied by the intensity at the center of the forward PSF. The two loss functions are

$$\mathcal{L}_{asym} = \frac{SR_f \; I_{f,center}}{SR_b}, \qquad \mathcal{L}_{uni} = \frac{P_{f,total} \; I_{f,center}}{P_{b,total}}, \tag{S7}$$

where $SR_f$, $SR_b$, $P_{f,total}$ and $P_{b,total}$ are the Strehl ratios and total focal-plane powers in the two directions, and $I_{f,center}$ is the intensity at the center of the forward PSF.

The gradient-descent optimization is run for a prespecified number of iterations. In each iteration, the gradient of the loss function, computed using the fully differentiable wave-propagation pipeline described above, is used to update the widths of the nanopillars constituting the doublet. Since each optic of the asymmetric imaging doublet has a diameter of 1 cm and contains a total of 25 million scatterers, four-fold symmetry is imposed to reduce the memory requirement.

## Supplementary Note 3. Details of the asymmetric imaging doublet fabrication and characterization

A 300 µm double-side-polished silicon wafer is cleaned and coated with positive photoresist (AZ 1505). The resist layer is then patterned with direct-write lithography (Heidelberg DWL 66+) and developed in commercial developer (AZ 726). Afterwards, the photoresist pattern is transferred into the bulk silicon and etched to a depth of 3 µm by deep reactive ion etching (SPTS DRIE). Finally, the photoresist residue is stripped away with oxygen plasma. The setup to measure the PSFs of the meta-doublet consists of a tunable quantum cascade laser (Daylight Solutions MIRcat mid-infrared laser) as the source, set to a wavelength of 4 µm. A relay system with 3x magnification is built to relay and magnify the focal-plane intensities of the meta-doublet onto the camera (FLIR A6780 MWIR) sensor, housed ~5 cm from the camera casing. The pixel size of the sensor is 15 x 15 µm$^2$ and the resolution is 640 x 512 pixels. The relay system comprises a 2.5 cm diameter, 2.5 cm focal length Ge plano-convex lens (Thorlabs) and a 5 cm diameter, 7.5 cm focal length $CaF_2$ plano-convex lens (Edmund Optics) as the objective and tube lens, respectively. During the PSF measurements, the meta-optics forming the doublet are aligned along the optical path and separated by a distance of 2 mm. An exposure time of 0.25 ms is used to record the focal-plane intensities.

## Supplementary Note 4. FWHM of the experimentally measured PSF

To determine the full width at half maximum (FWHM) of the measured focal spot, we take a horizontal slice of the experimentally measured PSF in the forward direction at $z$ = 9.5 mm and fit it to a Gaussian function. Supplementary Fig. 2 presents the camera-captured data points in the central 0.4 mm of the PSF along with the Gaussian fit. From this fit, the FWHM of the captured spot is 31 µm. Given that the relay system used in the PSF measurements has a magnification of 3x, this corresponds to a spot FWHM of 10.3 µm at the focal plane of the doublet. For comparison, a diffraction-limited optic with the doublet's numerical aperture of approximately 0.45 (1 cm diameter, 1 cm focal length) would produce a focal spot with an FWHM of approximately 0.51λ/NA ≈ 4.6 µm at λ = 4 µm. The measured spot is thus roughly twice the diffraction-limited width; the fabricated doublet is not diffraction limited, in line with the simulated forward Strehl ratio of 0.13 reported in the main text.

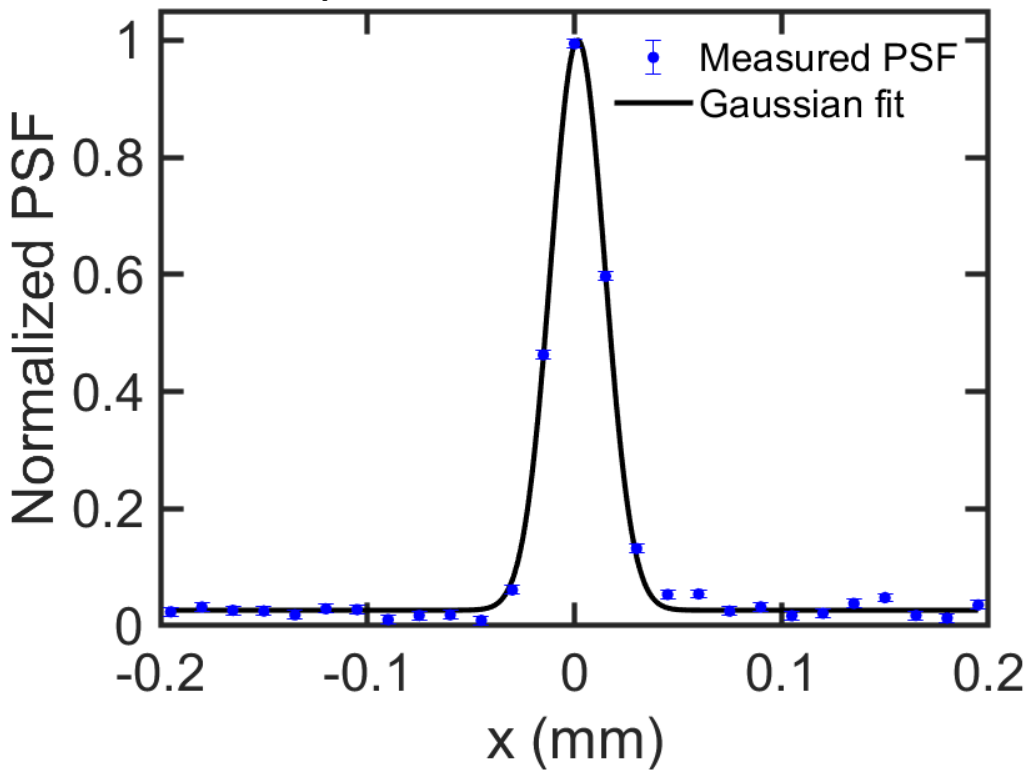

**Supplementary Figure 2:** Horizontal slice of the measured PSF of the asymmetric imaging doublet in the forward direction, fitted to a Gaussian function. Only the central 0.4 mm is shown for ease of visualization.

## Supplementary Note 5. Unidirectional imaging doublet

Here we design a doublet that, in addition to preferentially focusing a normally incident plane wave, also transmits significantly greater power in the forward direction than in the backward direction. In order to allow for greater design flexibility while ensuring that the optimization is computationally feasible, we design a smaller, 2 mm diameter doublet with a 0.5 mm spacing between the meta-optics. The design process and the loss function used during the optimization of the unidirectional doublet are described in Supplementary Note 2.

Supplementary Fig. 3(a) shows the feature maps of the optimized meta-optics, while Supplementary Fig. 3(b) presents the PSFs of the doublet in the forward and backward directions at a distance of 2 mm from it. For ease of visualization, both PSFs are normalized to their own maximum and only the central 0.4 x 0.4 $mm^2$ region is displayed. As designed, the doublet produces a focal spot in the forward direction and a random intensity pattern in the backward direction. The Strehl ratios of the doublet in the forward and backward directions are 0.73 and $5 \times 10^{-3}$, respectively. More importantly, the total focal-plane power in the forward direction, given by the PSF integrated over the focal-plane area, is approximately 158 times greater than that in the backward direction. This validates the unidirectional focusing performance of the designed doublet.

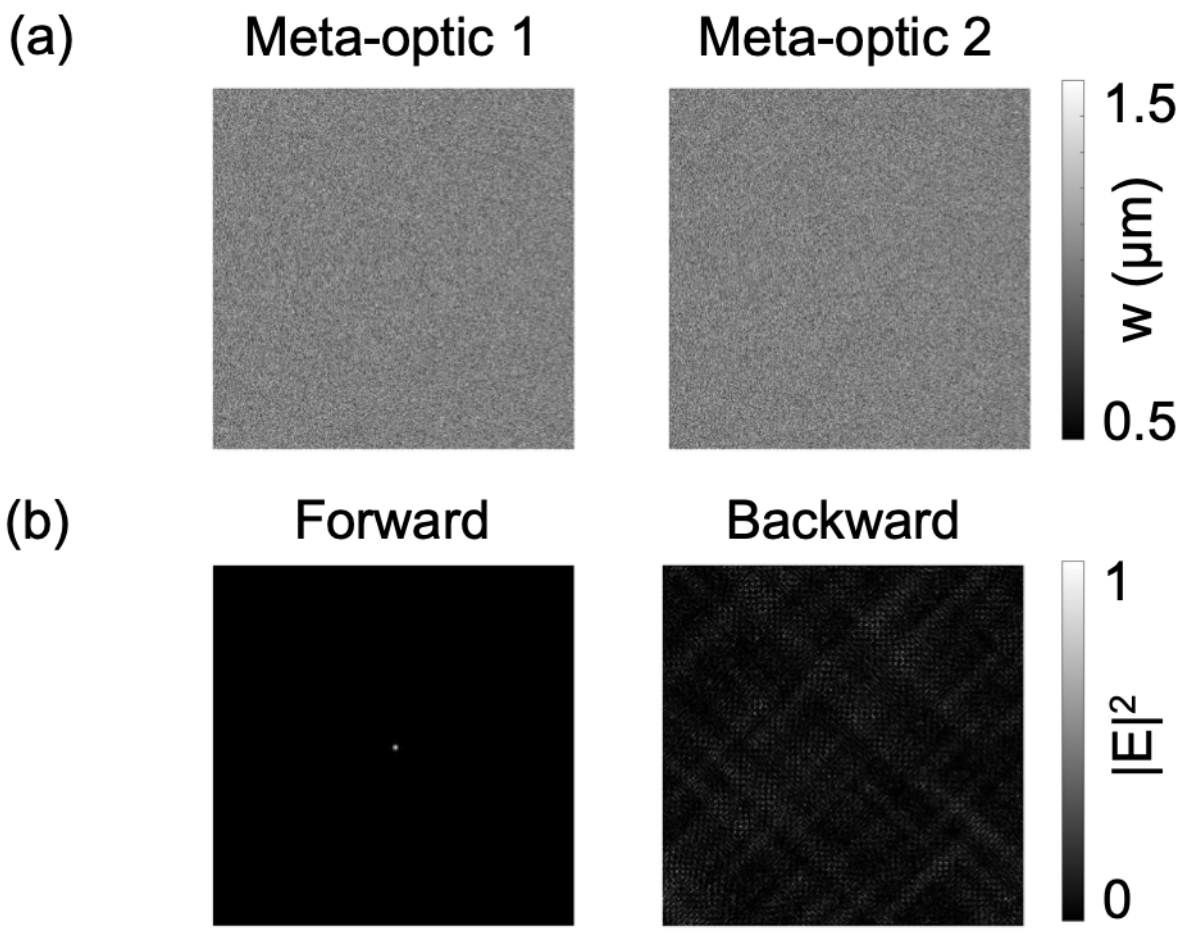

**Supplementary Figure 3:** (a) Feature maps of the meta-optics constituting the unidirectional imaging doublet. (b) Simulated PSFs of the doublet in the forward and backward directions.

**Supplementary Note 6. Working principle of the asymmetric imaging doublet**

To illustrate the working principle of our asymmetric imaging doublet, we present its PSFs and those of its constituent meta-optics at different distances along the optic axis in Supplementary Fig. 4. For ease of visualization, the PSFs are normalized to their own maximum and only the central 0.4 x 0.4 $mm^2$ area is displayed. We observe that meta-optic 1 produces a focal spot at a distance of 1.2 cm while meta-optic 2 produces a focal spot at a distance of 1 cm. The combined system produces a focal spot in the forward direction and a random intensity pattern in the backward direction at a distance of 1 cm from the trailing optic. Two additional focal spots are observed in the backward direction at distances of 8 and 12 mm from the trailing optic. Their respective peak intensities are 26 and 32 times smaller than that of the focal spot in the forward direction.

These observations suggest that the asymmetric imaging doublet can be viewed as a combination of two weakly modulating lenses with complementary focal lengths of 1 and 1.2 cm. The focal lengths of the two optics coincide only in the forward direction, thus producing a focal spot at a distance of 1 cm from the trailing optic. While this does not happen in the backward direction, the two optics are able to independently produce faint focal spots at distances of 8 and 12 mm. We note that the two meta-optics forming the doublet can also be considered as randomized hyperboloid lenses, where the level of randomness is optimized to achieve high focusing asymmetry without sacrificing forward-direction focusing performance.

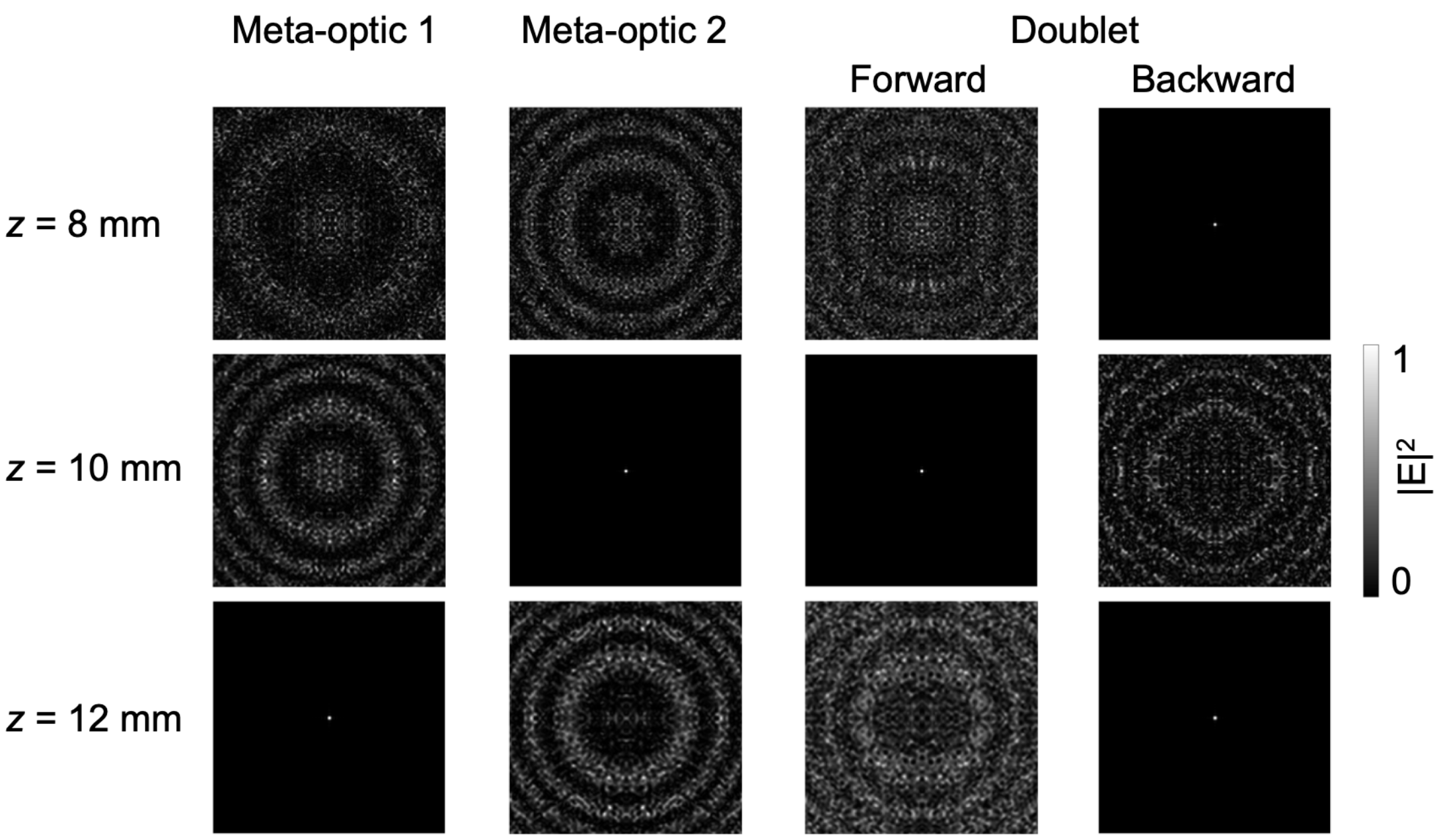

**Supplementary Figure 4:** Simulated PSFs of the asymmetric imaging doublet and of its constituent meta-optics at different distances *z* from the trailing meta-optic.

## Supplementary Note 7. Imaging simulations

To illustrate the asymmetric imaging capability of the doublets designed in this work, we perform imaging simulations using a phone-captured picture of a tree [Supplementary Fig. 5(a)]. The scene is assumed to be monochromatic and spatially incoherent, with the grayscale levels representing intensity values at a wavelength of 4 μm, so that the image produced by a given doublet in a particular direction is the two-dimensional convolution of the tree image with the doublet's simulated PSF in that direction. To reduce memory requirements, the tree image and the PSFs of the asymmetric imaging doublet (having an original size of 5000 x 5000 pixels) are resized to 1000 x 1000 pixels.

Supplementary Figs. 5(b) and 5(c) show the images obtained for the asymmetric imaging and unidirectional imaging doublets, respectively. To clearly show the power contrast between the imaging directions, both the forward and backward images of a given doublet are normalized to the maximum intensity of the corresponding forward-direction image. For the asymmetric imaging doublet, we observe that, despite being significantly blurred compared to the original image, the forward image preserves several of its features, in contrast to the backward-direction image. The total power transmitted in the backward direction is only slightly lower than that in the forward direction, as expected for a design whose loss function targets focusing asymmetry rather than power asymmetry. The unidirectional imaging doublet, on the other hand, achieves a significantly higher power contrast between the two imaging directions. Furthermore, the forward image for this doublet is much closer to the original scene, indicating superior imaging performance compared to the asymmetric imaging doublet. This is quantified by the forward Strehl ratios of 0.73 and 0.13 for the unidirectional and asymmetric imaging doublets, respectively. We note that these simulations use the on-axis PSFs at the design wavelength and assume spatially incoherent illumination; they are numerical illustrations of the doublets' response and are not experimental measurements.

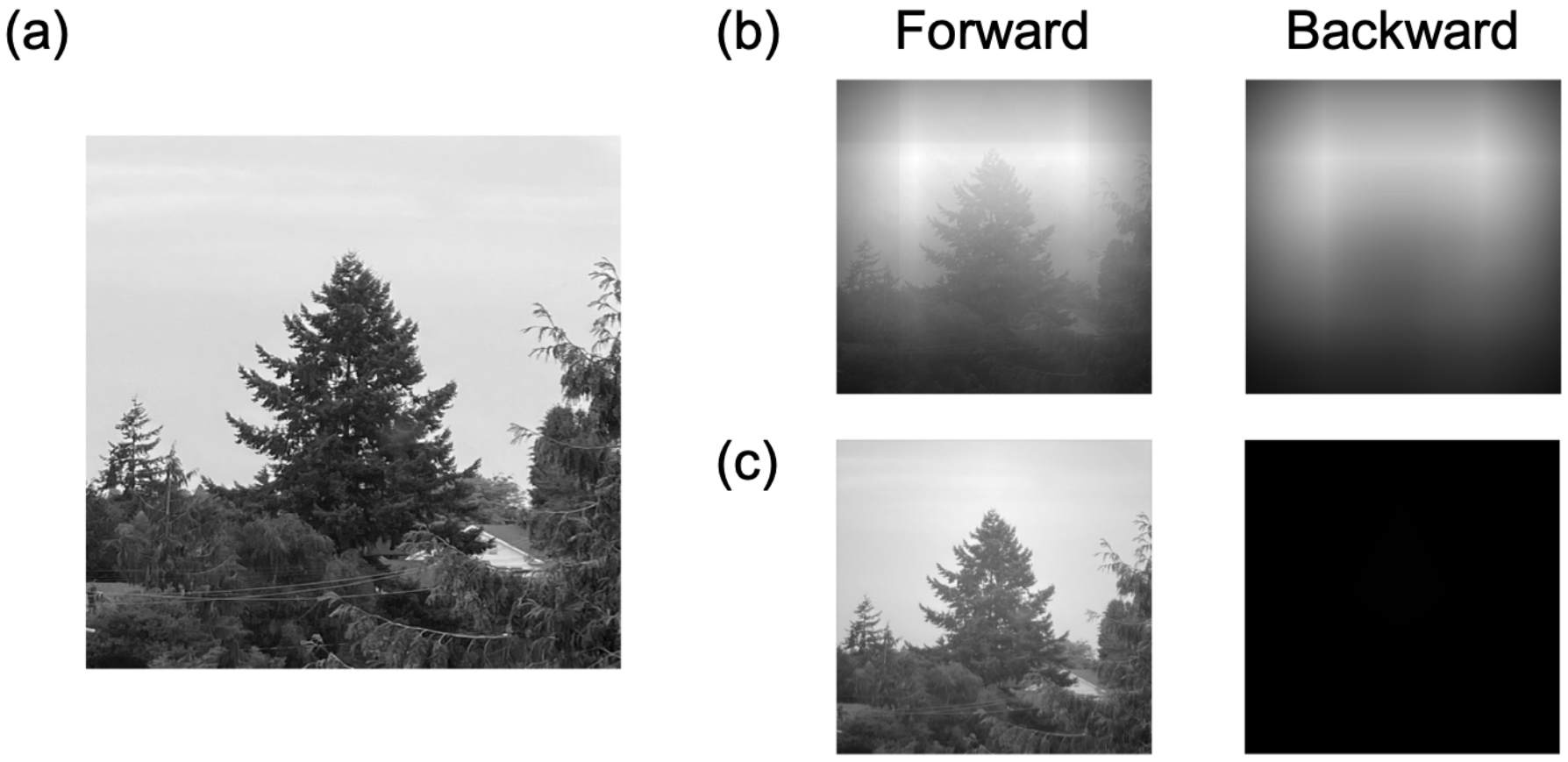

**Supplementary Figure 5:** (a) Grayscale image of a tree. Simulated images of the tree obtained using the calculated PSFs of the asymmetric imaging and unidirectional imaging doublets are shown in (b) and (c), respectively.